\documentstyle[preprint,revtex]{aps}
\begin{document}
%
\font\fortssbx=cmssbx10 scaled \magstep2
\hbox{ $\vcenter{\special{insert $disk1:[pheno.tex.inputs]uwlogo.imp}}$
\hskip.2in $\vcenter{\fortssbx University of Wisconsin - Madison}$ }
\begin{flushright}
\parbox{1.35in}{{\bf MAD/PH/701}\\{\bf IFT-P.014/92}\\hep-ph/9210277
\\October 1992}
\end{flushright}\par
\begin{title}
Threshold Effects on Heavy Quark Production in $\gamma\gamma$ Interactions
\end{title}
\author{O.\ J.\ P.\ \'Eboli$^1$, M.\ C.\ Gonzalez-Garcia$^2$,
F.\ Halzen$^2$ and\\  S.\ F.\ Novaes$^3$}
\begin{instit}
$^1$ Instituto de F\'{\i}sica,  Universidade
de S\~ao Paulo, \\
Caixa Postal 20516,  CEP 01498 S\~ao Paulo, Brazil.  \\
$^2$ Physics Department, University of Wisconsin, Madison, WI 53706,  USA \\
$^3$ Instituto de F\'{\i}sica Te\'orica, Universidade  Estadual Paulista, \\
Rua Pamplona 145, CEP 01405-900 S\~ao Paulo, Brazil.
\end{instit}

\begin{abstract}
The exchange of gluons between heavy quarks produced in $e^+e^-$ interactions
results in an enhancement of their production near threshold. We study QCD
threshold effects in $\gamma\gamma$ collisions. The results are relevant to
heavy quark production by beamstrahlung and laser back-scattering in future
linear collider experiments. Detailed predictions for top,
bottom and charm production are presented.
\end{abstract}


\newpage
\section{Introduction}

The QCD threshold enhancement of heavy quark production in $e^+e^-$
\cite{thr,fad:kho} and hadronic collisions  \cite{fad:kho:z} has been
profusely studied. In this paper we analyze this effect in
$\gamma\gamma$ collisions and we find a significant enhancement of top
production at future linear colliders. We consider two possibilities
for the sources of photons in an $e^+e^-$ machine: beamstrahlung
and laser back-scattering.

In studying the prospects for the commissioning of future $e^+e^-$
linear colliders \cite{lincol}, it has become clear that their physics
exploitation is inevitably affected by the fact that very dense electron and
positron bunches are also a very luminous source of photons. The strong
electromagnetic fields associated with the high charge density in such bunches
subject particles to very strong accelerating forces just prior to or during
the collision. As a result photons are radiated. This is known as beamstrahlung
\cite{nob,bd,jw}. The photon luminosity generated by beamstrahlung depends on
the characteristics of the beams, in particular on their transverse shape,
the length of the bunches, the number of electrons per bunch, and the
nominal beam energy. The desired photon luminosity can, in fact, be
achieved by tuning these parameters. We will focus our attention on the
design for the $500$ GeV collider NLC \cite{thesis,pal}, which is the
set G of parameters of Ref. \cite{pal}, and occasionally illustrate how results
change for different beam profiles and increased energy.

Beamstrahlung photons have a relatively soft spectrum. Hard photons can be
obtained by laser back-scattering. Here intense $\gamma$ beams are generated
by backward Compton scattering of soft photons from a laser of a few eV energy
\cite{las0}. The luminosity distribution over the $\gamma\gamma$ invariant
mass is broad and contains an abundant number of very energetic photons. The
angular spread from the Compton collision is small compared to the intrinsic
spread of the original electron beam and, therefore, the hard photon beam has
approximately the same cross-sectional area as the original electron beam.

The enhanced two-photon luminosity, whether from beamstrahlung or
laser back-scattering origin, is the source of a large number of $q \bar{q}$
pairs via two distinct mechanisms. The quarks can be generated by a direct
photon process, where the photons couple directly to charged quarks.
Alternatively, photons can interact via their quark and gluon constituents
\cite{witten}. This is referred to as a ``resolved" photon process. The
interaction of high energy photons via their quark or gluon structure leads to
the abundant production of hadron secondaries, thus giving rise to an
underlying event which gives the once clean $e^+e^-$ event the appearance of a
hadron collider interaction \cite{Drees}. Similarly the production of heavy
quarks by the two-photon process sprays the interaction region with a blizzard
of charm and beauty quarks and their associated prompt leptons \cite{ML}.
Two-photon processes also provide unique physics opportunities such as the
enhanced production of top quarks \cite{ML}. We revisit this problem paying
particular attention to  QCD enhancement of the threshold production of the
top quark in the  $\gamma\gamma$ process.

The layout of this paper is as follows. In Sec. \ref{heavy} we briefly
review the main features of heavy quark production by photons. In Sec.
\ref{distrib} we exhibit explicit expressions for the differential luminosities
${dL_{ij}}/{dz}$ for different sources of photons and partons. The
implementation of the QCD corrections for heavy quark production near threshold
are discussed in Sec. \ref{thres}. Section \ref{res} contains our results
and finally we summarize our conclusions in Sec. \ref{con}.

\section{Heavy quark production in $\gamma\gamma$ collisions}
\label{heavy}

The production of heavy quarks in $\gamma\gamma$ collisions can proceed
either by direct photons or by ``resolved'' photons. ``Resolved'' photons
produce heavy quark pairs via their quark and gluon constituents, which are
described in terms of the structure function of the partons in the photon
\cite{witten}. At tree level, there are four distinct contributions to heavy
quark pair production:
\begin{equation}
\begin{array}{l}
\gamma +\gamma \rightarrow Q\bar Q \\
\gamma +\gamma (g) \rightarrow Q\bar Q \\
\gamma(g) +\gamma(g) \rightarrow Q\bar Q \\
\gamma (q) +\gamma(\bar q) \rightarrow Q\bar Q \\
\end{array}
\label{Born}
\end{equation}
where $\gamma(g)$ and $\gamma(q)$ denotes a gluon or a quark component of the
photon respectively.  The expressions for these cross sections are well known
and can be found elsewhere \cite{Barger}.

The total cross section is obtained by folding the elementary cross section for
the processes (\ref{Born}) with the photon luminosity.
\begin{equation}
\sigma (e^+e^-\rightarrow \gamma\gamma\rightarrow i + j \rightarrow Q\bar Q)
(s)=\int_{z_{min}}^{z_{max}} dz~\frac{dL_{ij}}{dz}
\hat\sigma  (i + j \rightarrow Q\bar Q) (\hat s=z^2 s)
\end{equation}
Here $z^2= \tau = \hat{s}/s$, where $s$ is the total $e^+e^-$ CM energy squared
and $\hat{s}$ the $ij$ pair CM energy squared, and $dL_{ij}/dz$ stands
for the differential luminosity of the partons $i$ and $j$.

In order to obtain the total cross section, we must fix the characteristic
scales of the coupling constants and structure functions. We evaluate all
photon structure functions at the scale $Q^2 = \hat s/4$. The running strong
coupling constant is determined by the renormalization group
equation
\begin{equation}
\frac{d\alpha_s(Q^2)}{d\ln Q^2}=-b_0\alpha_s^2-b_1\alpha_s^3
+O(\alpha_s^3) \; ,
\label{running}
\end{equation}
with
\begin{equation}
b_0 = \frac{33-2N_f}{12\pi} \,\,\,\,\,\,\,\,\,
b_1 = \frac{153-19N_f}{24\pi^2} \;
\end{equation}
where $N_f$ is the number of active flavors. For tree level cross sections we
will use the first order ($b_1=0$) solution \begin{equation}
\alpha_s (Q^2) =\frac{12 \pi}{(33-2N_f)\ln(Q^2/\Lambda_n^2)} .
\label{alfas}
\end{equation}
At second order we will solve Eq. (\ref{running}) numerically. Flavor
thresholds are incorporated by choosing an appropriate value for $\Lambda_n$
which guarantees that $\alpha_s$ is continuous through the thresholds
$Q^2=m_i^2$ for $i=c,b,t$. Different values of $\Lambda_4$ will be chosen
corresponding to the different parametrizations of the photon structure
functions. Finally, we employ a running electromagnetic coupling, which in our
energy range is well described by
\begin{equation}
\alpha_{em}=\frac{1}{128-\frac{40}{9\pi}\ln(\sqrt{\hat s}/M_Z)} \; .
\end{equation}

\section{Distribution Functions}
\label{distrib}

The interpenetration of the dense electron and positron bunches in
future $e^+ e^-$ colliders generates strong accelerations on the
electrons and  positrons near the interaction point. This acceleration
gives rise to abundant bremsstrahlung. This phenomenon is known as
beamstrahlung \cite{nob,bd,jw}, and the distribution function of photons
created this way can be written  in the form
\begin{equation}
F_{\gamma/e}^B (x,b) = F^{(-)}_{\gamma/e}(x,b) \, \Theta(x_c - x) +
F^{(+)}_{\gamma/e}(x,b) \, \Theta(x - x_c)  \; .
\label{F:x}
\end{equation}
Here $x$ is the fraction of the beam energy carried by the photon, $b$
is the  impact parameter of the produced $\gamma$, and $x_c$ separates
low and high photon-energy regions  where different approximations to
$F_{\gamma/e}^B $ are used. The  distribution $F^{(-)}_{\gamma/e}$
adequate for small and intermediate values of $x$  is given by
\cite{bd,thesis}
\begin{eqnarray}
F^{(-)}_{\gamma/e}(x,b) \simeq && \frac{CK}{\Upsilon^{1/3}}
\left[ \frac{1 + (1-x)^2}{x^{2/3} (1-x)^{1/3}}  \right] \nonumber \\
&& \times \left \{ 1 + \frac{1}{6C \Upsilon^{2/3}}
\left(\frac{x}{1-x}\right)^{2/3}   \exp \left[ \frac{2}{3\Upsilon}
\frac{x}{(1-x)}  \right] \right\}^{-1} \; ,
\label{F:x0}
\end{eqnarray}
where $C = - Ai^\prime(0) = 0.2588$, and $Ai(x)$ is the Airy's function.
On the other hand, for large values of $x$, we have
\begin{equation}
F^{(+)}_{\gamma/e}(x,b) \simeq \frac{K}{2\sqrt{\pi}\Upsilon^{1/2}}
\left[ \frac{1 - x (1-x)}{x^{1/2}(1-x)^{1/2}}  \right]
\exp \left[ - \frac{2}{3\Upsilon} \frac{x}{(1-x)} \right] \; .
\label{F:x1}
\end{equation}

The value $x_c$ in Eq. (\ref{F:x}) is such that $F^B_{\gamma/e}$ is continuous
at $x=x_c$, i.e. $F^{(-)}_{\gamma/e} (x_c,b) =  F^{(+)}_{\gamma/e}(x_c,b)$. The
value of $x_c$ depends on the machine design, e.g. $x_c \simeq 0.48$ for the
original design for NLC. The dimensionless quantities $K$ and $\Upsilon$ are
defined as
\begin{eqnarray}
K & \equiv &  2\sqrt{3}\alpha \frac{\sigma_z E_\perp }{m} \; , \\ \nonumber
\Upsilon & \equiv & \frac{p E_\perp}{m^3} \; ,
\label{k:u}
\end{eqnarray}
where $m$ and $p$ are the electron mass and momentum, and $E_\perp$ is   the
transverse electric field inside a uniform elliptical bunch of dimensions
$l_{x,y} = 2 \sigma_{x,y}$ and $l_z = 2\sqrt{3}\sigma_z$,
\begin{equation}
E_\perp = \frac{N\alpha}{2\sqrt{3} (\sigma_x + \sigma_y)\sigma_z}
\left( \frac{b_x^2}{\sigma_x^2} + \frac{b_y^2}{\sigma_y^2}  \right)^{1/2}
\label{E}
\end{equation}
where $N$ in the number of particles in the bunch. For the original NLC  design
the value of these parameters is $\sigma_{x}=1.7\times 10^{-5}$ cm,
$\sigma_{y}=6.5\times 10^{-7}$ cm, $\sigma_{z}=0.011$ cm, and $N=1.67\times
10^{10}$.  We also study the effect of tuning to round beams by  choosing
$\sigma_{x(y)}=3.3\times10^{-6} cm$. For this case $x_c \simeq 0.64$ .

Notice that $F_{\gamma/e}^B (x,b)$  depends on the impact parameter through
$K$ and $\Upsilon$. In $\gamma\gamma$ collisions we should average  over the
impact parameters in order to obtain the actual photon-photon   luminosity
\begin{equation}
\frac{dL_{\gamma\gamma}^B}{dz} = 2 z
\int \frac{d^2 b}{4\pi\sigma_x\sigma_y} \int^1_{z^2} \frac{dx}{x}
F_{\gamma/e}^B (x,b) F_{\gamma/e}^B (z^2/x,b)  \; .
\label{b:eg}
\end{equation}
Therefore the necessity to average over the impact parameter implies that we
can not  decompose the effect of beamstrahlung into photon structure functions
\cite{thesis}.

The photon luminosity of beamstrahlung is very sensitive to the
transverse shape of the beam \cite{bd}. The aspect ratio,
\[
G = \frac{\sigma_x + \sigma_y}{2\sqrt{\sigma_x \sigma_y}}
\]
provides a good measure of beamstrahlung, with large photon
luminosities associated with small values of $G$. For high photon luminosity
one tunes to round beams {\em i.e.\/} $G=1$.  For the NLC original design
$G\simeq 2.7 $ \cite{thesis,pal}.

Conventional bremsstrahlung of photons by electrons further contributes to the
photon luminosity. This can be computed in the lowest order approximation
using the well-known Weisz\"{a}cker-Williams distribution
\begin{equation}
F_{WW}(x,E_{max})=\frac{\alpha}{2\pi} \frac{1 + (1 - x)^2}{x}
\ln\left(\frac{E^2_{max}}{m_e^2}  \right) \; ,
\label{WW}
\end{equation}
where $E_{max}$ is the electron beam energy.  The total $\gamma$
distribution is obtained by adding $F_{WW}$ to the beamstrahlung distribution
function $F_{\gamma/e}^B$.

The logarithm in Eq. (\ref{WW}) arises from the integration over the momentum
squared ($p^2$) of the photon propagator up to the maximum value
$E_{max}^2=s/2$. When computing cross sections we fold this distribution with
an elementary cross section which is evaluated for on-shell photons. The
effective  photon approximation is  valid only in the kinematical
regime where the elementary cross section does not depend on $p^2$.
It overestimates the number of off-shell photons. In order to
avoid this, we introduce a cutoff $E_{max}=E_{cut}$ in the integration
over the photon propagator which guarantees that the effective
photon approximation is used only in the kinematic range where it is strictly
valid. $E_{max}$ will be in general process dependent: in direct
$\gamma\gamma$ we will use the transverse momentum of the heavy quark as
a cutoff, otherwise we choose $E_{max} = 1$ GeV.  This procedure makes
the evaluation of the luminosities and cross sections conservative.

Abundant large invariant mass photons can also be obtained by the
process of laser back-scattering. When a laser light is focused almost head to
head on an energetic electron or positron beam we obtain a  large quantity of
photons carrying a great amount of the fermion energy. The energy spectrum of
back-scattered laser photons is \cite{las}
\begin{equation}
F_{\gamma/e}^L (x,\xi) \equiv \frac{1}{\sigma_c} \frac{d\sigma_c}{dx} =
\frac{1}{D(\xi)} \left[ 1 - x + \frac{1}{1-x} - \frac{4x}{\xi (1-x)} +
\frac{4
x^2}{\xi^2 (1-x)^2}  \right] \; ,
\label{F:laser}
\end{equation}
where $\sigma_c$ is the total Compton cross section. For the photons going in
the direction of the initial electron, the fraction $x$ represents the ratio
between the scattered photon and the initial electron energy ($x = \omega/E$).
In Eq.(\ref{F:laser}), we defined
\begin{equation}
D(\xi) = \left(1 - \frac{4}{\xi} - \frac{8}{\xi^2}  \right) \ln (1 + \xi) +
\frac{1}{2} + \frac{8}{\xi} - \frac{1}{2(1 + \xi)^2} \; ,
\label{D}
\end{equation}
with
\begin{equation}
\xi \equiv \frac{4 E \omega_0}{m^2} \cos^2 \frac{\alpha_0}{2} \simeq
\frac{2 \sqrt{s} \omega_0}{m^2} \; ,
\label{xi}
\end{equation}
where $\omega_0$ is the laser photon energy and ($\alpha_0 \sim 0$) is the
electron-laser collision angle.  The maximum value of $x$ is
\begin{equation}
x_m = \frac{\omega_m}{E} =  \frac{\xi}{1 + \xi} \; .
\label{ym}
\end{equation}

{}From Eq. (\ref{F:laser}) we can see that the fraction of photons with energy
close to the maximum value grows with $E$ and $\omega_0$.
Usually, the choice of $\omega_0$ is such that it is not possible for the
back-scattered photon to interact with the laser and create $e^+e^-$ pairs,
otherwise the conversion of electrons to photons would be dramatically reduced.
In our numerical calculations, we assumed $\omega_0 \simeq 1.26$ eV  for the
NLC
which is below the threshold of $e^+e^-$ pair creation ($\omega_m \omega_0 <
m^2$). Thus for the NLC beams we have $\xi \simeq 4.8$, $D(\xi) \simeq 1.9$,
and
$x_m \simeq 0.83$. In this case, half or more of the scattered photons are
emitted inside a small angle ($\theta < 5 \times 10^{-6}$ rad) and are very
energetic ($\omega > 100$ GeV).

The $\gamma\gamma$ luminosity from laser back-scattering is then
\begin{equation}
\frac{d L_{\gamma\gamma}^L}{dz} = 2z
k^2  \int_{z^2/x_{m}}^{x_{m}} \frac{dx}{x}
F_{\gamma/e}^L (x,\xi)F_{\gamma/e}^L (z^2/x,\xi) \; ,
\label{l:eg}
\end{equation}
where the conversion coefficient $k$ represents the average number of
high energy photons per one electron. We assume $k=1$ in our calculations.

Figure (\ref{f:lumi}.a) contains the differential $\gamma\gamma$ luminosities.
Beamstrahlung luminosity is shown for two different aspect  ratios ($G$) of the
beam at the NLC energies. In order to show the  bremsstrahlung contribution, we
plotted the $\gamma\gamma$ luminosity for beamstrahlung with and without
considering the bremsstrahlung photons. The actual $\gamma\gamma$ luminosity
will be somewhat reduced because no $E_{max}$ cutoff was implemented  in the
bremsstrahlung contribution in this figure . It is interesting to notice the
very steep dependence of the luminosity on $z$. In this figure we have also
shown the differential luminosity, Eq. (\ref{l:eg}), for laser back-scattering.
This luminosity is roughly constant in most of the $z<x_m$ range, as a
consequence of the hard  photon spectrum.

The luminosities, shown in Figure (\ref{f:lumi}.a) are valid for  interactions
where the photon couples directly to the quarks. Interactions
initiated by ``resolved" photons are described in terms of structure function
of partons, quarks and gluons, inside the photon \cite{witten}. We define an
effective distribution of partons in the electron by  folding the photon
structure functions with the photon distribution in the  electrons,
\begin{equation}
F^{L,B}_{p/e}(x,Q^2)=\int_x^1 \frac{dy}{y}
F^{L,B}_{\gamma/e} (x) P^\gamma (x/y,Q^2) \; ,
\end{equation}
where $P^\gamma = q^\gamma$ $(G^\gamma)$ is the quark (gluon) structure
function. Here, we also add the bremsstrahlung photons to the beamstrahlung
ones, and in this case an additional integration over impact parameter must be
performed. For ``resolved" photons the natural cutoff on the  bremsstrahlung
contribution is of order $\Lambda_{QCD}$. We use  $E_{cut}=1$ GeV. Also, the
suppression of the parton content of  highly off-shell photons is not a problem
with this choice, since we evaluate the parton distributions at $Q^2=\hat s/4$
with $Q^2>E^2_{cut}$, which guarantees that we do not include highly off shell
photons.

Finally, we define the parton-parton luminosity for once and twice
``resolved'' photons as
\begin{equation}
\frac{d L_{ij}^{L,B}}{dz} = 2 N z
\int_{z^2/x}^1 \frac{dx}{x}
F^{L,B}_{i/e} (x)F^{L,B}_{j/e}(z^2/x) \; ,
\end{equation}
where $i=\gamma$, $j=g$ for once ``resolved''  luminosity, and
$i=j=g$  or $i=q$ and $j=\bar q$ for twice ``resolved'' luminosities.
The statistical factor $N$ assumes
the value $N=2$ for distinct partons ($i\ne j$) and $N=1$ for identical partons
($i=j$).

The structure functions for partons inside the photon, $q^\gamma(x,Q^2)$ and
$G^\gamma(x,Q^2)$, are obtained for a given value of $Q^2=Q_0^2$ by fitting the
experimental data \cite{AMY}.    The $Q^2$ evolution is obtained, as
usual, by solving an  inhomogeneous  set of Altarelli-Parisi equations
\cite{lac,dewitt}. Several parametrizations have been proposed in the
literature \cite{lac,dg,do}. They lead to  different predictions as a
consequence of the large uncertainties due to the small number of experimental
results. In particular  very different parametrizations for $G^\gamma(x,Q^2)$
can fit the data.  We will present predictions  for the parametrizations of
Drees-Grassie (DG)  \cite{dg} and Levy-Abramowicz-Charchula (LAC3) \cite{lac},
which are  respectively characterized by a soft and a hard gluon distribution.
We take  $\Lambda_4=0.4$ GeV for the DG parametrization of the photon structure
functions and $\Lambda_4=0.2$ GeV for the LAC parametrizations.

In Fig. (\ref{f:lumi}.b) we show the once ``resolved" $\gamma g$ luminosities
for back-scattered laser  photons and beamstrahlung for ``ribbon-like'' beams,
using DG and LAC3  parametrizations of the parton distributions.  This figure
illustrates well the different behavior of the distributions DG and LAC3: for
back-scattered photons, the LAC3 $\gamma g$ luminosity  is larger (smaller)
then the corresponding one for DG at large (small) $z$.

Fig. (\ref{f:lumi}.c) and Fig. (\ref{f:lumi}.d) show the twice ``resolved"
$gg$ and $q\bar q$ luminosities (summed over the light quark flavors).  LAC3
parametrization predicts  a twice ``resolved''  luminosity always larger than
the one obtained with the DG parametrization.

\section{Threshold Behavior}
\label{thres}

The exchange of gluons between associatively produced heavy quarks modifies
significantly their production cross section near the threshold. Moreover, for
a very heavy quark, like the top, non-perturbative QCD effects are small, and
the threshold behavior can be computed perturbatively \cite{bigi,non:per},
since the top-quark width acts as an infrared cutoff .  In this case, the
modifications of the cross section near threshold due to QCD can be calculated
in terms of a Coulomb-like interaction between $t$ and $\bar t$.

In $\gamma\gamma$ collisions the $t\bar t$ pair can be produced in either a
color singlet or an octet state, depending on the production mechanism. The
threshold interaction  between the $t$ and $\bar t$ can be described by an
attractive Coulomb-like potential
\begin{equation}
V_S (r) = - \frac{4}{3} \frac{\alpha_s}{r}
\label{Vsin}
\end{equation}
in the color singlet channel, and by a repulsive potential
\begin{equation}
V_8 (r) = \frac{1}{6} \frac{\alpha_s}{r}
\label{Voc}
\end{equation}
in the color octet state. Since the interaction is attractive in the singlet
channel, the formation of bound states by multiple gluon exchanges between the
$t$ and the $\bar t$ can in principle occur. However, if the top quark is
heavier than $\sim 140$ GeV, the formation time of the bound state by gluon
exchange is larger than the lifetime of toponium and the resonance structure
disappears \cite{bigi,exi:top}. These interactions nevertheless lead to a
significant modifications of the cross section near threshold. This mechanism
is analogous to the Coulomb rescattering in QED discussed by Sommerfeld
\cite{sommer} and Sakharov \cite{sakha}.

In the narrow width approximation, we can obtain the QCD effects near the
threshold replacing, in the tree-level cross sections, the usual threshold
factor
\begin{equation}
\beta=\sqrt{1-\frac{4m_t^2}{\hat s}}   \; ,
\end{equation}
by
\begin{equation}
\beta |\Psi_{S,8}(0)|^2=\beta \frac{X_{S,8}}{1-\exp(-X_{S,8})}\equiv \beta
R_{S,8} \; ,
\label{factor}
\end{equation}
where $\Psi_{S,8}(0)$ is the wave function at the origin and
\begin{equation}
X_S=\frac{4}{3}\frac{\pi\alpha_s}{\beta} \;\; ,\;\;\;\;\;\;\;\;\;\;
X_8=-\frac{1}{6}\frac{\pi\alpha_s}{\beta} \; ,
\end{equation}
for the color singlet state ($S$), and octet ($8$) channels respectively.

Equation (\ref{factor}) can be interpreted as the exponentiated version of the
first order QCD corrections near the threshold. The first term in its
expansion in powers of $\alpha_s$ coincides with the one-loop QCD corrections.
The expression (\ref{factor}) does not include the effects of bound  states
below threshold  \cite{fad:kho,fad:kho:z}. These states are confined into a
very small energy region and their contribution to the total cross section,
which  is obtained by integration over all CM energies, is rather small.
Furthermore, unlike the $e^+e^-$ machines, it is not possible to observe the
effect of bounds states in the cross section through the $t \bar t$ excitation
curve due to the smearing introduced by the parton distribution functions.

Near threshold ($\beta \rightarrow 0$), the cross section in the color
singlet channel is increased since $\beta$ is substituted by the
non-vanishing factor $4\pi\alpha_s/3$. On the other hand, the octet
channel cross section is exponentialy suppressed in this limit. Therefore, the
factors in Eq. (\ref{factor}) are large, especially in the color
singlet channel, and this gives rise to a substantial enhancement of the
production cross section.

When computing the $t\bar t$ cross sections we use $\alpha_s$ in Eq.
(\ref{alfas}). The tree level cross sections are evaluated at $Q^2=\hat s/4$
while the QCD enhancement are given by
\begin{equation}
Q^2=p_{top}^2=m_t\sqrt{E^2+\Gamma_t^2}+ \frac{E^4}{4} \; ,
\end{equation}
where $E=\sqrt{\hat s}-2m_t$. We thus include the effect of the finite top
width
$\Gamma_t \approx 175\: m_t^3/M_W^3$ MeV.

In $\gamma\gamma$ collisions we have four contributions to top
production [see Eq. (\ref{Born})]. In the direct $\gamma\gamma$
interactions, the $t\bar t$ pair is produced in a singlet state.
Therefore the elementary cross section must be replaced by
\begin{equation}
\sigma^{th}(\gamma\gamma\rightarrow t\bar t)
=\sigma_0(\gamma\gamma\rightarrow t\bar t) R_{S} \; ,
\end{equation}
where $\sigma_0$ is the Born cross section. In $\gamma(g) +\gamma$
collision the $t\bar t$ is produced in the color octet channel because the
gluon is a color octet. The same is true in $\gamma(q)+\gamma(\bar q)$
annihilation  where a gluon in exchanged in the s-channel. In these cases we
have
\begin{equation}
\sigma^{th} (q \bar q ~(\gamma g) \rightarrow t \bar t)=\sigma_0 (q \bar q~
(\gamma g) \rightarrow
t \bar t) R_{8} \; .
\end{equation}

In $\gamma(g) +\gamma(g)$ fusion the final state is a mixture of
color singlet and octet states in a ratio 2 to 5 given by the color factors.
Therefore, we are lead to
\begin{equation}
\sigma^{th}(gg\rightarrow t\bar t)=\sigma_0(g g\rightarrow t\bar t)
\left( \frac{2}{7}R_{S}+\frac{5}{7}R_{8}\right) \; .
\end{equation}
Since the enhancement in the singlet channel is much larger than the
suppression
in the octet channel the net correction to  $g g$ is positive.
For the sake of comparison we also include the cross
sections for top production in direct $e^+ e^-$ annihilation. $t \bar t$
pairs produced in this channel are in the color singlet state. Therefore
we will have
\begin{equation}
\sigma^{th}(e^+e^-\rightarrow t\bar t)=R_S\sigma_0(e^+ e^-\rightarrow t\bar t).
\end{equation}

The previous analysis, valid for nonrelativistic particles, cannot be applied
to charm and bottom production. In this case bound state effects play a
critical role and the computation of the QCD enhancement becomes
non-perturbative near threshold. Here, we will compute the full
$O(\alpha^2\alpha_s)+O(\alpha\alpha_s^2) +O(\alpha_s^3)$ inclusive cross
section, $\gamma\gamma\rightarrow Q\bar Q [g,q ,\bar q]$, \cite{Smith,Ellis}
in the modified $\overline {MS}$ scheme as defined in \cite{Ellis}. We use the
value of $\alpha_s$ obtained by solving Eq. (\ref{running}) at second order.
We will show the results for two different scales $Q^2=m_i^2$ and $Q^2=4
m_i^2$, $i=c,b$. This procedure does not incorporate bound state effects but
should nevertheless represent an adequate estimate of the effect of the
threshold enhancement. The results will indicate that these corrections are
small relative to the uncertainty associated with the charm and bottom quark
masses. Again, we include the tree level and one-loop cross sections for charm
and bottom production in direct $e^+ e^-$ annihilation. The one-loop cross
section is given by \cite{Appelquist}
\begin{equation}
\sigma^{1-loop}_{e^+e^-}=\left [1 + \frac{4}{3} \, \alpha_s f(\beta)\right ]
\sigma^0_{e^+e^-} \; .
\label{1st}
\end{equation}
The function $f(\beta)$ \cite{Schwinger} is rather complicated involving
several Spence functions. Schwinger \cite{Schwinger} has constructed the
interpolating formula
\begin{equation}
f(x)=\frac{\pi}{2x}-\frac{3+x}{4}\left(\frac{\pi}{2}-\frac{3}{4\pi}\right)
\end{equation}
which agrees with the exact result to $1\%$ in the interval of interest.

\section{Results}
\label{res}

We are now ready to perform a full computation of heavy quark production in
$\gamma\gamma$ interaction including direct and ``resolved" photons and
incorporating QCD corrections near threshold. In Table \ref{tablet} we list the
production cross sections for top assuming  $m_{top}=120$ GeV and $\sqrt{s} =
500$ GeV. Contributions from different subprocesses are shown separately, with
and without threshold factors included for the sake of comparison. Results are
shown for beamstrahlung, laser back-scattering, and  direct $e^+e^-$
production.

As pointed out in Ref. \cite{Drees}, in the case of top production
the contribution of ``resolved" photons to the total $\gamma\gamma$ cross
section is small  as a result of the suppression of their luminosity at high
values of $x$, as can be seen from Figs. (1). Even for the LAC3
parametrization, characterized by a very hard gluon spectrum,
the contribution is at most $3\%$ for $\sqrt{s}=500$ GeV.
Since the direct singlet channel dominates, the threshold effect results
in a significant enhancement of the total cross section. This enhancement is
roughly a factor $2$ for beamstrahlung and more  than $50\%$ for laser
back-scattering.

In Fig. (\ref{f:inmass}) we show the invariant mass distribution of the $t\bar
t$ pair. The modifications  due to threshold effects are larger for small
invariant  masses, corresponding to $t\bar t$ pair production near threshold.
This explains why the QCD corrections are larger in $\gamma\gamma$  than in
$e^+e^-$ production. For the same reason the correction is small for laser
back-scattering  where the luminosity at low $x$ is suppressed.
Despite the corrections look big far from threshold we have checked that at
least $93\%$ of the effect in the total cross section comes from the region of
invariant mass less than $m_{top}$ above threshold.

The dependence on the top mass and on the collider energy is shown in  Fig.
(\ref{f:sigtop}). As expected, the QCD corrections increase  slightly with the
mass and decrease with the CM energy. As pointed out  in Ref. \cite{ML}
beamstrahlung, for round beams, can give a substantial  contribution to $t\bar
t$ production. The threshold corrections make this contribution even larger. At
$\sqrt{s}= 500$ GeV the two photon contributions is at most $10\%$ for the
``ribbon-like''  design. However, for a circular beam more than $50\%$ of the
$t\bar t$  pairs with $m_{top}<110$ GeV are produced in two photon collisions.
Since $\gamma\gamma$ cross section increases with energy while  $e^+e^-$ one
decreases, the two photon contributions are much more important  at $1$ TeV.
However top quarks produced by beamstrahlung photons preferentially populate
the low $p_T$ region and so do the prompt leptons from their decay [see Figs.
(\ref{f:pt}.a) and (\ref{f:pt}.b)]. In this case their signature suffers from
a large background from $b$ and $c$ produced both in direct $e^+e^-$
annihilation and in two photon processes.

The advantage of photon interactions is more dramatic for laser
back-scattering.
We first notice that at $\sqrt{s}=500$ $(1000)$ GeV, for $m_{top}< 130$ $(250)$
GeV, a ``$\gamma\gamma$ collider" can produce more $t\bar t$ pairs than the
corresponding $e^+e^-$ collider . The background from $c$ and $b$ quarks can be
efficiently suppressed because it is concentrated at lower $p_T$-values
than the one from the top signal. Furthermore, the separation of the signal
from the background is easier in this case than for direct $e^+e^-$ production
[compare Figs. (\ref{f:pt}.a) and (\ref{f:pt}.b)].

The cross sections at $\sqrt{s}=500$ GeV for inclusive charm and bottom
production are listed in Tables \ref{tablec} and \ref{tableb}.  At order
$\alpha^3$, with $\alpha=\alpha_{em}$ or $\alpha_s$, there are four new
contributions to the inclusive cross section apart from those of Eq.
(\ref{Born}):  $\gamma \gamma(q[\bar q])\rightarrow Q\bar Q q [\bar q] $ and
$\gamma(g) \gamma(q[\bar q])\rightarrow Q\bar Q  q [\bar q] $.  Note that
negative corrections only appear because we have separated the leading
order parton diagram from the QCD corrections; they should be added
\cite{Ellis}. The cross sections depend upon various factors, like the quark
mass, factorization scale, $\Lambda$, and the choice of parametrization of
the photon structure functions. Although the corrections to individual
channels can be large, the net modification of the total yield of heavy quarks
is smaller as a result of cancellation between opposite behaviors of the
different channels.  Even the sign of the correction depends on the quark
mass, the scale of  the coupling constant, the photon spectrum and the
parametrization of the photon structure functions.

We show results for two extreme values of the masses $m_c=1.35$ and $1.86$ GeV,
and $m_b=4.5$ and $5.2$ GeV.  The strong dependence on the mass makes
predictions relatively imprecise. In particular, the results are extremely
sensitive to the charm mass (by a factor of $\simeq 2$), while for bottom the
uncertainty is of the order of 40\%. This sensitivity to the heavy quark mass
was observed before in hadronic collisions and in $ep$ collisions \cite{Ellis}.

The dominant contribution to the total cross section comes from once
``resolved" $\gamma g$ process, due to the large $\gamma g$ luminosity. In
order to analyze the dependence on the parametrization of the structure
functions we evaluated the cross sections for DG and LAC3 parametrizations. In
contrast with the top case, the DG cross sections can be larger than those
computed with LAC3 structure functions. This is a result of a smaller
$\Lambda_4$-value and a harder gluon spectrum in the LAC3 parametrization. In
fact, we have checked that LAC1 indeed gives a  5--10 times bigger result for
the once ``resolved" process since parametrizations with softer spectra gives
rise to larger cross sections.   Moreover, uncertainty in the cross section due
to the structure functions is smaller for bottom production than for charm.

Finally, In order to estimate the size of higher order QCD corrections, we
computed  cross section for two different factorization scales: $Q^2=m_i^2$ and
$Q^2=4m_i^2$, with $i=c,b$  respectively. For charm production the  results
vary as much as $50\%$, while for bottom the variations are of the order of
$20\%$.

Despite the large values of the cross sections for charm and bottom,
most of them are produced at very low transverse momentum as shown in Fig.
(\ref{f:pt}). Therefore they will be hard to observe. If we impose a
transverse momentum cut on the prompt lepton of $10$ GeV, all cross
sections are reduced to less than $5$ pb. The main contribution to
large $p_T$ comes from the direct $\gamma\gamma$  process.

\section{Conclusions}
\label{con}

In this paper we have studied the QCD threshold effects on  heavy quark
production in $\gamma\gamma$ collisions. We have consistently taken into
account production by direct and ``resolved" photons.
We also studied how the cross section depends upon several factors
like quark mass, factorization scale, and choice of the structure functions.

Top quarks are predominantly produced in the direct $\gamma\gamma$ channel. In
this case, the $t\bar t$ pair is produced  through a color singlet channel and
the threshold  effect results in a substantial enhancement of the total cross
section.  At $\sqrt{s}=500$ GeV the enhancement is a factor $2$ for
beamstrahlung and  more than $1.5$ for laser back-scattering. For a given
collider energy,  it will increase with the top mass, while for a given mass it
decreases with the collider energy.

For charm and bottom the contributions due to ``resolved" photons are dominant,
mainly via the once ``resolved" $\gamma+\gamma (g)$  process.  The effect of
the correction is always smaller than the uncertainty due to the  choice of the
bottom and charm masses.

\acknowledgments

This work is supported in part by the U.S.~Department of Energy under
contract No.~DE-AC02-76ER00881, and in part by the University of
Wisconsin Research Committee with funds granted by the Wisconsin Alumni
Research Foundation. Further support was given by Conselho Nacional de
Desenvolvimento Cient\'\i fico e Tecnol\'ogico (CNPq) and Funda\c c\~ao
de Amparo \`a Pesquisa do Estado de S\~ao Paulo (FAPESP).

We would like to thank J. Smith for providing us with the QCD corrections
for the charm and bottom production and S. Keller for supplying us with
the photon distributions functions. We also thank D. Zeppenfeld for useful
discussions.
Two of us (O.J.P.E. and S.F.N.) are very grateful to the Institute for
Elementary Particle Physics Research, University of  Wisconsin --
Madison for their kind hospitality during the initial step of this work.

\figure{Differential luminosities in $\gamma\gamma$ collisions at
$\sqrt{s}=500$ GeV. In (a) the direct photon luminosities are shown for
back-scattered laser photons (solid line), beamstrahlung photons for
``ribbon-like'' beam  with G=2.7 (dashed lines) and  round beam (dotted lines).
The lower dashed and dotted lines show the luminosity without including  the
bremsstrahlung photons and the upper ones include the bremsstrahlung  photons
according to the EPA distribution of Eq. (\ref{WW}). In (b) the  once
``resolved" $\gamma g$ luminosities are shown for back-scattered laser  photons
(solid and dotted lines) and beamstrahlung for ``ribbon-like'' beam  (dashed
and dot-dashed lines). The solid and dashed lines correspond to the  DG
parametrization of partons inside the photons and the dotted and dot-dashed
ones correspond to the LAC3 parametrization. (c) and (d) show the twice
``resolved" $gg$  and $q\bar q$ luminosities for the same  cases as (b).
\label{f:lumi}}

\figure{Invariant mass distributions of the $t\bar t$ pairs produced  in
$\gamma\gamma$ collisions for $m_t=120$ GeV. Figure (a) correspond to
back-scattered laser photons while (b) and (c) correspond to   beamstrahlung
photons for G=2.7 and G=1 respectively. In all cases the solid (dotted) lines
show the distributions with (without)  the threshold factors.\label{f:inmass}}

\figure{Total $t\bar t$ production cross section as a function of $m_t$ for
$\sqrt{s}=500$ GeV (a) and $\sqrt{s}=1$ TeV (b). Solid lines are the cross
sections for direct $e^+e^-$ production, dashed lines for $\gamma\gamma$
production with back-scattered laser photons, and dotted (dot-dashed) lines
for $\gamma\gamma$ production with beamstrahlung photons with G=2.7 (G=1).
In all cases the upper (lower) lines show the cross section with (without)
the threshold factors. \label{f:sigtop}}

\figure{Transverse momentum distribution of the lepton produced in the
heavy quark decay. Dashed, dotted, and dot-dashed lines
correspond respectively to charm, bottom and top produced in
$\gamma\gamma$ process with back-scattered laser photons (a)
and beamstrahlung photons for G=2.7 (b)
and G=1 (c). Solid lines correspond to annihilation production
of the heavy quarks according to labels in the figure.
Quark masses were assumed to be $m_c=1.86$ GeV, $m_b=5.2$ GeV,
and $m_t=120$ GeV.\label{f:pt}}

\begin{table}
\begin{center}
\begin{displaymath}\small
\begin{array}{||c|c|c|c|c|c|c||}
\hline
\hline
\mbox{\bf process} & \multicolumn{6}{c||}
{\mbox{\bf cross section (pb)}} \\
\hline
\hline
e^+e^-   &  \multicolumn{3}{c|} { 0.70 } &
\multicolumn{3}{c||} {0.94  }\\
\hline
\hline
\multicolumn{7}{c||}{ } \\[-0.95cm] \mbox{photon-photon}  &
\multicolumn{2}{c|} {\mbox{laser}} & \multicolumn{2}{c|}  {\mbox{G=2.7}} &
\multicolumn{2}{c||} {\mbox{G=1}} \\[-0.1cm] \hline
\gamma+\gamma    & 0.74  & 1.2   & 9.0\times 10^{-3} & 1.8\times 10^{-2}
& 0.18  & 0.36  \\
\hline
\gamma+\gamma(g) &
\begin{array}{c} 4.2\times 10^{-3}\\[-0.3cm] 1.7\times 10^{-2} \end{array} &
\begin{array}{c} 3.8\times 10^{-3}\\[-0.3cm] 1.5\times 10^{-2}\end{array}   &
\begin{array}{c} 1.2\times 10^{-5}\\[-0.3cm] 4.8\times 10^{-5}\end{array}   &
\begin{array}{c} 1.1\times 10^{-5}\\[-0.3cm] 4.4\times 10^{-5}\end{array}   &
\begin{array}{c} 3.5\times 10^{-4}\\[-0.3cm] 1.4\times 10^{-3}\end{array}   &
\begin{array}{c} 3.1\times 10^{-4}\\[-0.3cm] 1.3\times 10^{-3}\end{array}   \\
\hline
\gamma(g)+\gamma(g) &
\begin{array}{c} 5.4\times 10^{-7}\\[-0.3cm] 9.7\times 10^{-6}\end{array}   &
\begin{array}{c} 7.1\times 10^{-7}\\[-0.3cm] 1.2\times 10^{-5}\end{array}   &
\begin{array}{c} 1.4\times 10^{-9}\\[-0.3cm] 2.4\times 10^{-8}\end{array}   &
\begin{array}{c} 1.8\times 10^{-9}\\[-0.3cm] 3.1\times 10^{-8}\end{array}   &
\begin{array}{c} 3.1\times 10^{-8}\\[-0.3cm] 5.5\times 10^{-7}\end{array}   &
\begin{array}{c} 4.2\times 10^{-8}\\[-0.3cm] 7.2\times 10^{-7}\end{array}   \\
\hline
\gamma(q)+\gamma(\bar{q}) &
\begin{array}{c} 2.5\times 10^{-4}\\[-0.3cm] 2.8\times 10^{-4}\end{array}   &
\begin{array}{c} 2.2\times 10^{-4}\\[-0.3cm] 2.5\times 10^{-4}\end{array}   &
\begin{array}{c} 8.8\times 10^{-7}\\[-0.3cm] 9.7\times 10^{-7}\end{array}   &
\begin{array}{c} 7.8\times 10^{-7}\\[-0.3cm] 8.7\times 10^{-7}\end{array}   &
\begin{array}{c} 2.6\times 10^{-5}\\[-0.3cm] 2.9\times 10^{-5}\end{array}   &
\begin{array}{c} 2.3\times 10^{-5}\\[-0.3cm] 2.6\times 10^{-5}\end{array}   \\
\hline
\hline
\end{array}
\end{displaymath}
\end{center}
\caption{Cross sections for $t\bar t$ production at $\sqrt{s}=500$ GeV
for $m_t=120$ GeV.
The first row corresponds to $e^+e^-$ annihilation production and
the others correspond to photon-photon production.
For each process the left (right)
column is the cross section without (with) the threshold factors.
We separate the different contributions to the photon-photon cross sections
from direct photons, $\gamma + \gamma$, once ``resolved"
gluon-photon fusion, $\gamma + \gamma(g)$, and twice ``resolved''
gluon fusion, $\gamma(g) + \gamma(g)$, and  $\gamma(q) + \gamma(\bar q)$
annihilation. For ``resolved" photon processes the  upper number is the
cross section with DG parametrization and the lower one is the cross section
with LAC3 parametrization.}
\label{tablet}
\end{table}
\begin{table}
\begin{center}
\begin{displaymath}\footnotesize
\begin{array}{||c|c|c|c|c|c|c|c||}\hline
\hline
\mbox{\bf process} & \mbox{\bf $m_c$ (GeV)} & \multicolumn{6}{c||}
{\mbox{\bf cross section (nb)}} \\
\hline
\hline
e^+e^-   & 1.35-1.86  &  \multicolumn{3}{c|}{0.77\times 10^{-3}} &
\multicolumn{3}{c||}{0.83\times 10^{-3}-0.87\times 10^{-3} } \\
\hline
\hline
\multicolumn{8}{c||}{ } \\[-0.8cm]
\begin{array} {c} \mbox{photon-}\\ [-0.3cm]  \mbox{photon}\end{array} &
   &  \multicolumn{2}{c|} {\mbox{laser}} & \multicolumn{2}{c|}
{\mbox{G=2.7}} & \multicolumn{2}{c||} {\mbox{G=1}} \\[-0.1cm]
\hline
\gamma\gamma &1.35   & 0.165  & 0.365-0.305  & 12.1  & 19.4-17.3  & 47.0  &
75.1-66.9   \\
\cline{2-8}
& 1.86  &0.138   &0.264-0.231  & 6.68  &10.1-9.22   & 26.8 & 40.1-36.8  \\
\hline
\gamma\gamma(g) &1.35 &
\begin{array} {c} 266-170 \\ [-0.3cm] 96.9-71.1\end{array} &
\begin{array} {c}374-151 \\ [-0.3cm]  201-69.8\end{array} &
\begin{array} {c}87.7-55.9 \\ [-0.3cm] 66.1-49.7\end{array} &
\begin{array} {c} 104-56.0\\ [-0.3cm] 92.2-47.0\end{array} &
\begin{array} {c}629-401 \\ [-0.3cm]  408-299  \end{array} &
\begin{array} {c}771-390 \\ [-0.3cm]  600-282\end{array} \\
\cline{2-8}
 & 1.86 &
\begin{array} {c} 120-82.7 \\ [-0.3cm]  50.4-38.5\end{array} &
\begin{array} {c} 151-86.8\\ [-0.3cm]   82.4-37.6\end{array} &
\begin{array} {c} 33.3-23.0\\ [-0.3cm]   31.2-23.8\end{array} &
\begin{array} {c} 38.0-25.5\\ [-0.3cm] 37.5-23.6\end{array} &
\begin{array} {c} 249-172\\ [-0.3cm] 194-148 \end{array} &
\begin{array} {c} 290-188\\ [-0.3cm] 246-145 \end{array} \\
\hline
\begin{array}{c}
\gamma\gamma(q)\\ + \end{array}
 &1.35 &
\begin{array} {c} 0.0-0.0 \\ [-0.3cm] 0.0-0.0\end{array} &
\begin{array} {c} 26.4-0.513\\ [-0.3cm]  16.6-1.24\end{array} &
\begin{array} {c}0.0-0.0 \\ [-0.3cm] 0.0-0.0\end{array} &
\begin{array} {c} 9.18- (-1.13)\\ [-0.3cm] 6.34- (-0.674)\end{array} &
\begin{array} {c}0.0-0.0 \\ [-0.3cm]  0.0-0.0  \end{array} &
\begin{array} {c} 70.7- (-5.21) \\ [-0.3cm]  48.2- (-2.26)\end{array} \\
\cline{2-8}
\gamma\gamma(\bar q)
 & 1.86 &
\begin{array} {c} 0.0-0.0 \\ [-0.3cm]  0.0-0.0\end{array} &
\begin{array} {c} 8.11- (-0.095)\\ [-0.3cm]   5.98-0.300\end{array} &
\begin{array} {c} 0.0-0.0\\ [-0.3cm]   0.0-0.0\end{array} &
\begin{array} {c} 2.47- (-0.524)\\ [-0.3cm] 1.95- (-0.383)\end{array} &
\begin{array} {c} 0.0-0.0\\ [-0.3cm] 0.0-0.0 \end{array} &
\begin{array} {c} 19.8- (-2.80)\\ [-0.3cm] 15.6- (-1.75) \end{array} \\
\hline
\gamma(g)\gamma(g) &1.35 &
\begin{array} {c} 38.2-15.5\\ [-0.3cm]  48.7-26.2\end{array} &
\begin{array} {c} 51.2-18.1\\ [-0.3cm]  95.7-31.7\end{array} &
\begin{array} {c} 5.44-2.21\\ [-0.3cm]  18.2-9.77\end{array} &
\begin{array} {c} 6.47-2.84\\ [-0.3cm]  24.3-11.0\end{array} &
\begin{array} {c} 49.7-20.2\\ [-0.3cm]  127-68.5\end{array} &
\begin{array} {c} 60.4-25.1\\ [-0.3cm]  185-77.1\end{array} \\
\cline{2-8}
 & 1.86 &
\begin{array} {c}  11.2-5.34\\ [-0.3cm]  20.5-11.9\end{array} &
\begin{array} {c} 14.2-6.86\\ [-0.3cm]  31.5-13.9\end{array} &
\begin{array} {c} 1.31-0.623\\ [-0.3cm]  6.54-3.81\end{array} &
\begin{array} {c} 1.55-0.855\\ [-0.3cm] 7.80-4.47\end{array} &
\begin{array} {c} 12.7-6.05\\ [-0.3cm]  47.8-27.8\end{array} &
\begin{array} {c} 15.2-8.12\\ [-0.3cm]  60.0-32.3\end{array} \\
\hline
\gamma(q)\gamma(\bar{q}) &1.35 &
\begin{array} {c} 0.681-0.276\\ [-0.3cm]  0.304-0.164\end{array} &
\begin{array} {c} 0.225-0.173\\ [-0.3cm]  0.148-0.117\end{array} &
\begin{array} {c} 0.685-0.278\\ [-0.3cm]  0.330-0.178\end{array} &
\begin{array} {c} 0.321-0.235\\ [-0.3cm]  0.165-0.139\end{array} &
\begin{array} {c} 3.89-1.58\\ [-0.3cm]  1.89-1.02\end{array} &
\begin{array} {c} 1.72-1.27\\ [-0.3cm]  0.932-0.779\end{array} \\
\cline{2-8}
 &1.86 &
\begin{array} {c} 0.322-0.153\\ [-0.3cm]  0.159-0.092\end{array} &
\begin{array} {c} 0.143-0.116\\ [-0.3cm]  0.083-0.070\end{array} &
\begin{array} {c} 0.205-0.097\\ [-0.3cm]  0.144-0.084\end{array} &
\begin{array} {c} 0.108-0.085\\ [-0.3cm]  0.078-0.069\end{array} &
\begin{array} {c} 1.28-0.607\\ [-0.3cm]  0.853-0.497\end{array} &
\begin{array} {c} 0.650-0.515\\ [-0.3cm]  0.459-0.400\end{array} \\
\hline
\begin{array}{c}
\gamma(g)\gamma(q)\\ + \end{array}
 &1.35 &
\begin{array} {c} 0.0-0.0 \\ [-0.3cm] 0.0-0.0\end{array} &
\begin{array} {c} 15.2-0.344\\ [-0.3cm]  14.6-1.66\end{array} &
\begin{array} {c}0.0-0.0 \\ [-0.3cm] 0.0-0.0\end{array} &
\begin{array} {c} 1.75- (-0.145)\\ [-0.3cm] 2.81- 0.043\end{array} &
\begin{array} {c}0.0-0.0 \\ [-0.3cm]  0.0-0.0  \end{array} &
\begin{array} {c} 17.2- (-0.762) \\ [-0.3cm]  23.7-1.14\end{array} \\
\cline{2-8}
\gamma(g)\gamma(\bar q)
 & 1.86 &
\begin{array} {c} 0.0-0.0 \\ [-0.3cm]  0.0-0.0\end{array} &
\begin{array} {c} 3.15- (-0.059)\\ [-0.3cm]   4.03-0.352\end{array} &
\begin{array} {c} 0.0-0.0\\ [-0.3cm]   0.0-0.0\end{array} &
\begin{array} {c} 0.312- (-0.056)\\ [-0.3cm] 0.659- (-0.041)\end{array} &
\begin{array} {c} 0.0-0.0\\ [-0.3cm] 0.0-0.0 \end{array} &
\begin{array} {c} 3.22- (-0.039)\\ [-0.3cm] 5.86- (-0.062) \end{array} \\
\hline
\mbox{total} &1.35 &
\begin{array} {c}  305-186\\[-0.3cm] 146-97.6  \end{array} &
\begin{array} {c}  467-170\\ [-0.3cm]328-105 \end{array} &
\begin{array} {c}  106-70.5 \\ [-0.3cm]96.7-71.7  \end{array} &
\begin{array} {c}  141-75.1\\ [-0.3cm]145-74.8  \end{array} &
\begin{array} {c}  730-470\\ [-0.3cm] 584-416 \end{array} &
\begin{array} {c}  496-477\\[-0.3cm] 750-426  \end{array} \\
\cline{2-8}
 &1.86 &
\begin{array} {c} 132-88.3\\[-0.3cm]71.2-50.6   \end{array} &
\begin{array} {c} 177-93.9\\ [-0.3cm]124-52.5  \end{array} &
\begin{array} {c} 41.5-30.4\\ [-0.3cm]44.6-34.4  \end{array} &
\begin{array} {c} 52.5-35.1\\ [-0.3cm]58.1-36.9  \end{array} &
\begin{array} {c} 290-206\\ [-0.3cm] 269-203 \end{array} &
\begin{array} {c} 369-231\\ [-0.3cm]368-213  \end{array} \\
\hline
\end{array}
\end{displaymath}
\end{center}
\caption{Same as Table I for charm production. The cross sections
are listed for two quark masses. For each process the two right (left)
numbers are the tree level (one loop) cross sections. In each case
the left number is for $Q^2=m_c^2$ and the
right one for $Q^2=4m_c^2$.}
\label{tablec}
\end{table}

\begin{table}
\begin{center}
\begin{displaymath}\footnotesize
\begin{array}{||c|c|c|c|c|c|c|c||}
\hline
\hline
\mbox{\bf process} &\mbox{\bf $m_b$ (GeV)} & \multicolumn{6}{c||}
{\mbox{\bf cross section (pb)}} \\
\hline
\hline
e^+e^-   & 4.5-5.2  &  \multicolumn{3}{c|}{0.41}  &
\multicolumn{3}{c||}{0.44} \\
\hline
\hline
\multicolumn{8}{c||}{ } \\[-0.85cm]
\begin{array} {c} \mbox{photon-}\\ [-0.3cm]  \mbox{photon}\end{array} &
   &  \multicolumn{2}{c|} {\mbox{laser}} & \multicolumn{2}{c|}
{\mbox{G=2.7}} & \multicolumn{2}{c||} {\mbox{G=1}} \\[-0.1cm]
\hline
\gamma\gamma & 4.5 & 4.91  & 7.21-6.80 & 74.6  &101-96.3   & 332  & 445-425  \\
\cline{2-8}
&5.2  &4.43  &6.31-5.99  & 55.3 &73.9-70.9  &251  & 332-319   \\
\hline
\gamma\gamma(g) &4.5 &
\begin{array}{c}  1820-1440 \\[-0.3cm] 1590-1320  \end{array} &
\begin{array}{c}  2030-1600 \\[-0.3cm] 1860-1340  \end{array} &
\begin{array}{c}  371-293 \\[-0.3cm]  644-533\end{array} &
\begin{array} {c} 398-343 \\[-0.3cm]  675-574 \end{array} &
\begin{array}{c}  3030-2390  \\[-0.3cm]  4430-3670   \end{array} &
\begin{array} {c} 3260-2750 \\[-0.3cm]   4720-3890\end{array} \\
\cline{2-8}
 &5.2 &
\begin{array} {c}  1140-910\\[-0.3cm]  1090-915 \end{array} &
\begin{array} {c}  1250-1010\\[-0.3cm] 1240-933 \end{array} &
\begin{array} {c}  218-175\\[-0.3cm]  409-342 \end{array} &
\begin{array} {c}  233-205\\[-0.3cm] 426-372\end{array} &
\begin{array} {c}  1810-1450\\[-0.3cm] 2870-2400  \end{array} &
\begin{array} {c}  1940-1680\\[-0.3cm] 3020-2570  \end{array} \\
\hline
\begin{array}{c} \gamma\gamma(q) \\ + \end{array} &4.5 &
\begin{array}{c}  0.0-0.0 \\[-0.3cm] 0.0-0.0 \end{array} &
\begin{array}{c}  99.6- (-10.2) \\[-0.3cm] 93.7- (-5.23) \end{array} &
\begin{array}{c}  0.0-0.0 \\[-0.3cm]  0.0-0.0 \end{array} &
\begin{array} {c} 18.7- (-11.1) \\[-0.3cm]  17.5- (-10.9) \end{array} &
\begin{array}{c}  0.0-0.0 \\[-0.3cm]  0.0-0.0 \end{array} &
\begin{array} {c} 178- (-69.8) \\[-0.3cm]   168- (-66.8) \end{array} \\
\cline{2-8}
 \gamma\gamma(\bar q)   &5.2 &
\begin{array} {c}  0.0-0.0 \\[-0.3cm]  0.0-0.0 \end{array} &
\begin{array} {c}  61.1- (-7.63) \\[-0.3cm] 58.9- (-4.96) \end{array} &
\begin{array} {c}  0.0-0.0 \\[-0.3cm]  0.0-0.0 \end{array} &
\begin{array} {c}  10.2- (-7.27) \\[-0.3cm] 9.62- (-7.35) \end{array} &
\begin{array} {c}  0.0-0.0 \\[-0.3cm] 0.0-0.0 \end{array} &
\begin{array} {c}  101- (-47.0) \\[-0.3cm] 96.7- (-46.8) \end{array} \\
\hline
\gamma(g)\gamma(g) &4.5 &
\begin{array} {c} 287-179 \\[-0.3cm] 1310-898 \end{array} &
\begin{array}  {c} 331-241 \\[-0.3cm]  1510-1080 \end{array} &
\begin{array} {c}  21.7-13.5 \\[-0.3cm] 244-167 \end{array} &
\begin{array} {c}  25.1-19.6 \\[-0.3cm]   269-219 \end{array} &
\begin{array} {c} 242-151 \\[-0.3cm]   2080-1430 \end{array} &
\begin{array} {c} 278-213 \\[-0.3cm]  2300-1820 \end{array} \\
\cline{2-8}
 &5.2 &
\begin{array} {c}  156-99.8 \\[-0.3cm]  794-555 \end{array} &
\begin{array} {c}  178-135 \\[-0.3cm]  898-677 \end{array} &
\begin{array} {c}  10.8-6.94 \\[-0.3cm]   134-93.9 \end{array} &
\begin{array} {c}  12.5-10.1 \\[-0.3cm]   148-125 \end{array} &
\begin{array} {c}  124-79.6 \\[-0.3cm]  1180-826 \end{array} &
\begin{array} {c}  143-114 \\[-0.3cm]   1300-1070 \end{array} \\
\hline
\gamma(q)\gamma(\bar{q}) &4.5 &
\begin{array} {c}  24.1-15.0 \\[-0.3cm]   22.5-15.4 \end{array} &
\begin{array} {c}  14.5-13.0 \\[-0.3cm]   14.0-12.9 \end{array} &
\begin{array} {c}  10.7-6.68 \\[-0.3cm]  11.9-8.19 \end{array} &
\begin{array} {c}  6.63-6.02 \\[-0.3cm]  7.76-7.33 \end{array} &
\begin{array} {c}  73.8-46.0 \\[-0.3cm]  80.6-55.3 \end{array} &
\begin{array} {c}  45.1-40.9 \\[-0.3cm]  51.7-48.6 \end{array} \\
\cline{2-8}
 &5.2 &
\begin{array} {c}  15.7-10.1 \\[-0.3cm]   15.8-11.1 \end{array} &
\begin{array} {c}  9.64-8.79 \\[-0.3cm]   10.0-9.39 \end{array} &
\begin{array} {c}  6.66-4.27 \\[-0.3cm]   7.70-5.39 \end{array} &
\begin{array} {c}  4.20-3.88 \\[-0.3cm]   5.11-4.88 \end{array} &
\begin{array} {c}  46.6-29.9 \\[-0.3cm]   53.2-37.2 \end{array} &
\begin{array} {c}  29.0-26.8 \\[-0.3cm]   34.8-33.1 \end{array} \\
\hline
\begin{array}{c} \gamma(g)\gamma(q) \\ + \end{array} &4.5 &
\begin{array}{c}  0.0-0.0 \\[-0.3cm] 0.0-0.0 \end{array} &
\begin{array}{c}  48.9- (-11.8) \\[-0.3cm] 107- (-9.68) \end{array} &
\begin{array}{c}  0.0-0.0 \\[-0.3cm]  0.0-0.0 \end{array} &
\begin{array} {c} 2.54- (-2.25) \\[-0.3cm]  9.06- (-5.52) \end{array} &
\begin{array}{c}  0.0-0.0 \\[-0.3cm]  0.0-0.0 \end{array} &
\begin{array} {c} 33.2- (-19.9) \\[-0.3cm]   101- (-39.6) \end{array} \\
\cline{2-8}
 \gamma(g)\gamma(\bar q) &5.2 &
\begin{array} {c}  0.0-0.0 \\[-0.3cm]  0.0-0.0 \end{array} &
\begin{array} {c}  24.5- (-7.52) \\[-0.3cm] 57.3- (-8.14) \end{array} &
\begin{array} {c}  0.0-0.0 \\[-0.3cm]  0.0-0.0 \end{array} &
\begin{array} {c}  1.09- (-1.26) \\[-0.3cm] 4.18- (-3.38) \end{array} &
\begin{array} {c}  0.0-0.0 \\[-0.3cm] 0.0-0.0 \end{array} &
\begin{array} {c}  15.2- (-11.5) \\[-0.3cm] 49.2- (-25.5) \end{array} \\
\hline
\mbox{total} &4.5 &
\begin{array} {c} 2140-1640 \\[-0.3cm] 2930-2240  \end{array} &
\begin{array} {c} 2530-1840 \\[-0.3cm] 3590-2430  \end{array} &
\begin{array} {c} 478-388   \\[-0.3cm] 974-783    \end{array} &
\begin{array} {c} 552-452   \\[-0.3cm] 1080-880   \end{array} &
\begin{array} {c} 3680-2920 \\[-0.3cm] 6920-5490  \end{array} &
\begin{array} {c} 4240-3340 \\[-0.3cm] 7790-6070  \end{array} \\
\cline{2-8}
 &5.2 &
\begin{array} {c} 1320-1020 \\[-0.3cm] 1900-1490  \end{array} &
\begin{array} {c} 1530-1150 \\[-0.3cm]  2270-1610 \end{array} &
\begin{array} {c} 291-242   \\[-0.3cm]   606-497  \end{array} &
\begin{array} {c} 335-281   \\[-0.3cm]   667-562  \end{array} &
\begin{array} {c} 2230-1810 \\[-0.3cm] 4350-3510  \end{array} &
\begin{array} {c} 2560-2080 \\[-0.3cm] 4830-3920  \end{array} \\
\hline
\end{array}
\end{displaymath}
\end{center}
\caption{Same as Table II for bottom.}
\label{tableb}
\end{table}

\end{document}